\documentclass[sn-mathphys,Numbered]{sn-jnl}

\usepackage{graphicx}%
\usepackage{multirow}%
\usepackage{amsmath,amssymb,amsfonts}%
\usepackage{amsthm}%
\usepackage{mathrsfs}%
\usepackage[title]{appendix}%
\usepackage{xcolor}%
\usepackage{textcomp}%
\usepackage{manyfoot}%
\usepackage{booktabs}%
\usepackage{algorithm}%
\usepackage{algorithmicx}%
\usepackage{algpseudocode}%
\usepackage{listings}%
\DeclareMathOperator{\sech}{sech}

\theoremstyle{thmstyleone}%
\newtheorem{theorem}{Theorem}
\theoremstyle{thmstyletwo}%
\newtheorem{remark}{Remark}%

\theoremstyle{thmstylethree}%
\newtheorem{definition}{Definition}%

\begin{document}

\title[Article Title]{Witt type Realizations of $2$-D Cayley-Klein Algebras with non-zero curvatures}


\author*[1]{\fnm{Arindam} \sur{Chakraborty}}\email{arindam.chakraborty@heritageit.edu}



\affil*[1]{\orgdiv{Department of Physics}, \orgname{Heritage Institute of Technology}, \orgaddress{\street{Chowbaga Rd, Anandapur, Mundapara}, \city{Kolkata}, \postcode{700107}, \state{West Bengal}, \country{India}}}




\abstract{The article presents various Witt type vector field realizations of 2-D Cayley-Klein algebras with non-vanishing curvatures. The expressions of the vector fields involve Jacobi elliptic functions whose moduli are directly related to the parameters that appear in the corresponding matrix representation obtained from a bi-orthogonal set of vectors. First, the realizations are obtained with the values of the  moduli lying in the unit interval $(0, 1)$. The parameter of bi-orthogonality plays a crucial role in this context. Later, with the help of modular transformation, realizations involving arbitrary moduli have been obtained. }




\maketitle

\section{Introduction}\label{sec1}
The main objective of the present article is to investigate Witt type \cite{xiao17} realizations of the so called \textbf{Cayley-Klein (henceforth CK) Algebras} with non-zero curvatures. Based on Cayley's work and later developed as part of Klein's Erlangen Program \cite{book06, erlangen25} these algebras are found to have close relation with metric spaces of various geometric types. This is possible in the spirit of Klein's observation that behind any motion group there is a notion of geometric space and the idea leads to an unified description of geometric spaces. Apart from its significance in unifying geometries and contraction of Lie algebras \cite{sanju84, lord85, gromov90, herranz94, nieto97}, CK-algebras have been applied in various other fields of interest starting from geometric algebra \cite{daniel13, gunn17}, geometric flows \cite{benson20}, non-commutative geometry \cite{herranz14, herranz21},  quantum mechanics \cite{gromov90.1, gromov90.2, ball93, herranz19, panahi20}, relativity \cite{sila08} and cosmology \cite{mad93}, particle physics \cite{herranz21} to recent surge of interest in machine learning \cite{nielsen16, bi17}. 

We may recall the notion of representation of a Lie algebra in which it is imperative to construct matrices or differential operators (often called vector fields) that satisfy a desired set of commutation relations \cite{xiao17, de lucas20}. In the present case we first construct $2\times 2$ matrices representing the triplet  of generators of each of the CK-algebras with non-zero curvatures (elliptic, hyperbolic, co-hyperbolic, doubly hyperbolic) starting from a set of \textbf{bi-orthogonal pairs of vectors}. Similar vectors have been considered earlier by the present author  \cite{chakraborty20} in the context of non-Hermitian quantum mechanics. Next, we go for a vector fields representation of the said CK-algebras in terms of Jacobi's three elementary elliptic functions \cite{lawden89,akhiezer90}. In fact, an attempt to solve  a triplet of non-linear coupled ordinary differential equations has facilitated the process of obtaining the said elliptic functions when fitted with the relevant commutation relations of the normalized CK-algebra. The parameter of bi-orthogonality plays a crucial role in this process by relating itself with the \textbf{modulus} and \textbf{complimentary modulus} of the said elliptic functions. Initially, a variety of representations are obtained with elliptic functions whose moduli lie in the open unit interval $(0, 1)$. Later the so called \textbf{modular transformation} \cite{lawden89, khare04} has been taken into account and consequently the realizations involving arbitrary absolute values of the moduli have been obtained. Finally, the trigonometric and hyperbolic limit of the said vector fields have been considered. It has been observed that in some occasions it is possible to obtain non-trivial vector fields involving trigonometric/hyperbolic functions at the respective limits, whereas in some other cases taking such limit does not produce three distinct vector fields. There exists atleast one case where such a limiting procedure is not at all allowed and hence the vector fields involve functions which are purely elliptic in nature allowing no trigonometric or hyperbolic analogues. A brief discussion on Jacobi elliptic functions and modular transformation have been presented as an appendix.

It is interesting to note that the conventional expressions of Casimirs relating to all of the present realizations in terms of vector fields becomes zero, which is however not unusual since such Casimirs depend on representation involved therein. However, following \cite{thomas06} we consider a kind of intrinsic Casimir operators which are dependent on only the commutation relations among the generators, not on the specific representation under question. Here, Casimir is understood as a differential invariant whose Lie derivative is zero.

\section{Caley-Klein algebras and their normalized forms}\label{ck}
Following Sanjuan \cite{sanju84}, Lord \cite{lord85},  Gromov \cite{gromov90} and Nieto et. al. \cite{nieto97} the map $\mathcal{F}: \mathbb{R}^3\rightarrow \mathbb{R}^3(\mathbf{j})$ can be defined by the following actions
\begin{equation}
\mathcal{F}x_0^\star=x_0,\:\:\:
\mathcal{F}x_1^\star=j_1x_1,\:\: {\rm{and}}\:\:\mathcal{F}x_2^\star=j_1j_2x_2
\end{equation}
The nine possible CK-geometries can be realized on a sphere
 \begin{equation}
S_2(\mathbf{j})=x_0^2+j_1^2x_1^2+j_1^2j_2^2x_2^2=1.
\end{equation}
where, $\mathbf{j}=(j_1, j_2)$, $j_1=1, i, \iota_1$ and $j_2=1, i, \iota_2$ with $\iota_1^2=\iota_2^2=0; \iota_1\iota_2=\iota_2\iota_1\neq 0$.

Motivated from \cite{gromov90, nieto97}, let us consider $\mathfrak{so}_3(\mathbf{j})$ as the Lie algebra denoted by three generators $\{J, P_1,P_2\}$ with commutators 

\begin{equation}\label{CK}
[J, P_1]=P_2,\:\: [J, P_2]=-j_2^2P_1=-\kappa_2P_1,\:\: [P_1, P_2]=j_1^2J=\kappa_1J
\end{equation}

The following is a list of four possible normalized CK-algebras with non-vanishing curvatures $(\kappa_1\neq 0, \kappa_2\neq 0)$.

\begin{table}[h]
\caption{Table for normalized CK-algebra of non-vanishing curvatures $(\kappa_1, \kappa_2)\neq(0, 0)$}
\label{ta0}
	{\begin{tabular}{|c|c|c|c|} \toprule
			Type of CK algebra & $(\kappa_1, \kappa_2)$ & Normalized generators & Commutation relations\\\hline
			 &  & $\widetilde{P}_1=\kappa_1^{-\frac{1}{2}}P_1$   & $[\widetilde{J}, \widetilde{P}_1]=\widetilde{P}_2$\\
			Elliptic & $(1, 1)$ & $\widetilde{P}_2=(\kappa_1\kappa_2)^{-\frac{1}{2}}P_2$ & $[\widetilde{J}, \widetilde{P}_2]=-\widetilde{P}_1$\\
			& &$\widetilde{J}=\kappa_2^{-\frac{1}{2}}J$ & $[\widetilde{P}_1, \widetilde{P}_2]=\widetilde{J}$\\\hline
			 &  & $\widetilde{P}_1=-\kappa_1^{-\frac{1}{2}}P_1$ & $[\widetilde{J}, \widetilde{P}_1]=\widetilde{P}_2$\\
			Doubly-Hyperbolic/de-Sitter & $(-1, -1)$ & $\widetilde{P}_2=(\kappa_1\kappa_2)^{-\frac{1}{2}}P_2$ & $[\widetilde{J}, \widetilde{P}_2]=\widetilde{P}_1$\\
			& &$\widetilde{J}=-\kappa_2^{-\frac{1}{2}}J$ & $[\widetilde{P}_1, \widetilde{P}_2]=-\widetilde{J}$\\
			\hline
			 &  & $\widetilde{P}_1=i\kappa_1^{-\frac{1}{2}}P_1$ & $[\widetilde{J}, \widetilde{P}_1]=\widetilde{P}_2$\\
			Hyperbolic/Lobachevski & $(-1, 1)$ & $\widetilde{P}_2=i(\kappa_1\kappa_2)^{-\frac{1}{2}}P_2$ & $[\widetilde{J}, \widetilde{P}_2]=-\widetilde{P}_1$\\
			& & $\widetilde{J}=\kappa_2^{-\frac{1}{2}}J$ & $[\widetilde{P}_1, \widetilde{P}_2]=-\widetilde{J}$\\
			\hline
			 &  & $\widetilde{P}_1=-i\kappa_1^{-\frac{1}{2}}P_1$ & $[\widetilde{J}, \widetilde{P}_1]=\widetilde{P}_2$\\
			 Co-Hyperbolic/Anti-de-Sitter & $(1, -1)$ & $\widetilde{P}_2=i(\kappa_1\kappa_2)^{-\frac{1}{2}}P_2$ & $[\widetilde{J}, \widetilde{P}_2]=\widetilde{P}_1$\\
			& & $\widetilde{J}=-\kappa_2^{-\frac{1}{2}}J$ & $[\widetilde{P}_1, \widetilde{P}_2]=\widetilde{J}$\\
			\hline
		\end{tabular} }
\end{table}

The quadratic Casimirs for the above algebras are given by 
$\mathcal{C}_E=P_1^2+P_2^2+J^2$, $\mathcal{C}_{dS}=P_2^2-P_1^2-J^2$, $\mathcal{C}_{L}=P_2^2+P_1^2-J^2$ and $\mathcal{C}_{AdS}=-P_1^2+P_2^2+J^2$ respectively.

\section{Bi-orthogonal system and Cayley-Klein Algebras}\label{biortho}
\textbf{Definition}:Two pairs of vectors $\{\vert\phi_j\rangle : j=1, 2\}$ and $\vert\chi_j\rangle : j=1, 2$ in $\mathbb{C}^2$ are said to form a \textbf{bi-orthogonal set} relative to the inner-product $\langle\cdot\vert\cdot\rangle$ if $\langle\phi_j\vert\chi_k\rangle=\delta_{jk}$.

Let us consider $\vert v_j\rangle=\left(\begin{array}{c}
	c^{(1)}_j \\
	c^{(2)}_j
\end{array} \right)\in \mathbb{C}^2$ and the inner-product $\langle v_j\vert v_k\rangle=(c^{(1)}_j)^\star c^{(1)}_k+(c^{(2)}_j)^\star c^{(2)}_k$. The following theorem is immediate \cite{chakraborty20}.

\vspace{0.5cm}

\begin{theorem}
Given a pair of vectors $\{\vert v_j\rangle : j=1,2\}$ and an operator 
\begin{equation}\label{trans}
T=\cos\frac{\vartheta}{2}\mathbf{1}_{2\time 2}+2\cos\frac{\varphi}{2}\sin\frac{\vartheta}{2}\sigma_1-2\sin\frac{\varphi}{2}\sin\frac{\vartheta}{2}\sigma_2,
\end{equation}
 the set $\{\vert\phi_j\rangle= T \vert v_j\rangle: j=1, 2\}$ and $\{\vert\chi_j\rangle= (T^{-1})^\dagger\vert v_j\rangle: j=1, 2\}$ constitute bi-orthogonal system provided
 $\langle v_j\vert v_k\rangle=0$; $\{\sigma_m : m=1, 2, 3\}$ are \textbf{Pauli matrices}. 
\end{theorem}

\vspace{0.5cm}

Proof : Since $T=T^\dagger$, $\langle\phi_j\vert\chi_k\rangle=\langle v_j\vert T^\dagger (T^\dagger)^{-1}\vert v_k\rangle=\langle v_j\vert v_k\rangle$. This relation is valid for all $\vartheta$. Hence, follows the theorem.$\:\:\:\square$.

\vspace{0.5cm}

Acting $T$ and $(T^{-1})^\dagger$ on the orthogonal vectors $\{\vert q_j\rangle=2^{-\frac{1}{2}}\left(\begin{array}{c}
	1 \\
	(-1)^{j-1}
\end{array} \right) : j=1, 2\}$,   we get, for $\varphi=\pi$

\begin{eqnarray}
\{\vert\phi_j\rangle= T\vert q_j\rangle= 2^{-\frac{1}{2}}\left(\begin{array}{c}
	e^{-i(\frac{3}{2}-j\vartheta)} \\
	(-1)^{j-1}e^{i(\frac{3}{2}-j\vartheta)} 
\end{array} \right) : j=1, 2\}\:\:{\rm{and}}\nonumber\\
\{\vert\chi_j\rangle= (T^{-1})^\dagger\vert q_j\rangle= 2^{-\frac{1}{2}}\left(\begin{array}{c}
	e^{i(\frac{3}{2}-j\vartheta)} \\
	(-1)^{j-1}e^{-i(\frac{3}{2}-j\vartheta)} 
\end{array} \right) : j=1, 2\}
\end{eqnarray}

leading to the following generators 

\begin{equation}
\sigma_m^\gamma=\frac{i^{m+1}}{2}\sum_{j, k=1}^{2}\frac{c_{jk}^{(m)}}{\omega^{\delta_{m2}}}\vert \phi_j\rangle\langle \chi_k\vert : m=1, 2, 3
\end{equation}.
 Here, $\omega=\sqrt{1-\gamma^2}=\cos\theta$ and $c_{jk}^{(1)}=(-1)^j\delta_{jk}$ and $c_{jk}^{(3)}=(-1)^jc_{jk}^{(2)}=(1-(-1)^jc_{jk}^{(1)})$.
 In explicit forms,
\begin{equation}
\sigma_1^\gamma={\frac{1}{2}}\left(\begin{array}{cc}
	-i\gamma & 1\\
	1 & i\gamma
\end{array} \right),\:\: \sigma_2^\gamma={\frac{1}{2}}\left(\begin{array}{cc}
	0 & -i\\
	i & 0
\end{array} \right)\:\:{\rm{and}}\:\:\sigma_3^\gamma={\frac{1}{2}}\left(\begin{array}{cc}
	1 & i\gamma\\
	i\gamma & -1
\end{array} \right)
\end{equation}

Defining $\widetilde{J}=\frac{i}{\omega}\sigma^\gamma_1$, $\widetilde{P}_1=i\sigma^\gamma_2$ and $\widetilde{P}_2=-\frac{i}{\omega}\sigma^\gamma_3$, the following commutation relations are possible
\begin{equation}\label{ellipticCK}
[\widetilde{J}, \widetilde{P}_1]=\widetilde{P}_2,\:\:[\widetilde{J}, \widetilde{P}_2]=-\widetilde{P}_1,\:\:[\widetilde{P}_1, \widetilde{P}_2]=\widetilde{J}
\end{equation}
Equation-\ref{ellipticCK} represents the \textbf{elliptic} CK-algebra.

Making the following choices like
\begin{eqnarray}
\left\{\widetilde{J}=\frac{i}{\omega}\sigma^\gamma_1, \widetilde{P}_1=\sigma^\gamma_2, \widetilde{P}_2=-\frac{1}{\omega}\sigma^\gamma_3\right\}\nonumber\\
\left\{\widetilde{J}=-\frac{1}{\omega}\sigma^\gamma_1, \widetilde{P}_1=i\sigma^\gamma_2, \widetilde{P}_2=\frac{1}{\omega}\sigma^\gamma_3\right\}\nonumber\\
\left\{\widetilde{J}=\sigma^\gamma_2, \widetilde{P}_1=-\frac{1}{\omega}\sigma^\gamma_1, \widetilde{P}_2=\frac{i}{\omega}\sigma^\gamma_3\right\}
\end{eqnarray}
 one can
represent the sets of generators of \textbf{hyperbolic, co-hyperbolic} and \textbf{doubly hyperbolic} algebras respectively.

\vspace{0.5cm}

\section{Witt realizations of CK-algebras via elliptic functions}\label{witt}
\subsection{Realizations by elliptic functions of modulus $k$}\label{witt1}
\begin{definition}
The Witt realization \cite{xiao17} of an $n$-dimensional Lie algebra $\mathcal{L}$ with generators $\{T_j : j=1\cdots n\}$, involves meromorphic functions of one independent variable $u$ such that $\{T_j=f_j(u)\frac{d}{du} : j=1\cdots n\}$ along with the Lie bracket $[T_i, T_j]=[f_i\frac{df_j}{du}-f_j\frac{df_i}{du}]\frac{d}{du}$.
\end{definition}
 
The above mentioned CK-algebra may be represented in terms of vector fields i. e.; $\{\widetilde{J}, \widetilde{P}_1, \widetilde{P}_2\}\equiv\{f_1\frac{d}{dz}, f_3\frac{d}{dz}, f_2\frac{d}{dz}\}$ with the introduction of the following triplet of non-linear differential equations
\begin{equation}\label{nonlin}
{\rm{(a)}}\:\:\frac{df_1}{dz}+\lambda_1 f_2f_3=0\:\:\:
{\rm{(b)}}\:\:\frac{df_2}{dz}+\lambda_2 f_1f_3=0\:\:\:
{\rm{(c)}}\:\:\frac{df_3}{dz}+\lambda_3 f_2f_1=0,
\end{equation}

where, $\{\lambda_j:j=1, 2, 3\}\in\mathbb{R}\backslash\{0\}$. Let us choose $f_1=\frac{g_1}{\sqrt{\lambda_2}}, f_2=ig_2$ and $f_3=-i\frac{g_3}{\sqrt{\lambda_2}}$

Multiplying equation-\ref{nonlin}-(b) with $f_1$ and equation-\ref{nonlin}-(a) 
with $f_2$ and comparing the result with the relation $[\widetilde{J}, \widetilde{P}_2]=-\widetilde{P}_1$, we get
\begin{equation}\label{lamb1}
\lambda_1f_2^2-\lambda_2f_1^2=-1\:\:{\rm{or}}\:\:f_1=\pm\sqrt{\frac{1+\lambda_1f_2^2}{\lambda_2}}=\pm \frac{g_1}{\sqrt{\lambda_2}}
\end{equation}

Similar considerations with equation-\ref{nonlin}-(b) and equation-\ref{nonlin}-(c)
and using $[\widetilde{P}_1, \widetilde{P}_2]=\widetilde{J}$
\begin{eqnarray}\label{lamb2}
\lambda_2f_3^2-\lambda_3f_2^2=-1\:\:{\rm{or}}\:\:f_3=\pm i\sqrt{\frac{1-\lambda_3f_2^2}{\lambda_2}}=\pm i\frac{g_3}{\sqrt{\lambda_2}}
\end{eqnarray}

Putting all these results in equation-\ref{nonlin}-(b), we get 
\begin{eqnarray}\label{jacobi2}
\frac{dg_2}{dz}\pm\sqrt{(1-\lambda_1g_2^2)(1+\lambda_3g_2^2)}=0\nonumber\\
\frac{dg_2}{dz}=\pm\sqrt{(1-\lambda_1g_2^2)(1+\lambda_3g_2^2)}
\end{eqnarray}

Considering equation-\ref{nonlin}-(a) and equation-\ref{nonlin}-(c) and using $[\widetilde{J}, \widetilde{P}_1]=\widetilde{P}_2$

\begin{equation}\label{lamb3}
\lambda_1f_3^2-\lambda_3f_1^2=1\:\:{\rm{or}}\:\:f_3=\pm \sqrt{\frac{\lambda_2+\lambda_3g_1^2}{\lambda_1\lambda_2}}
\end{equation}

On the other hand, in view of equation-\ref{lamb1}, $f_2=\pm\sqrt{\frac{g_1^2-1}{\lambda_1}}$

 Now from equation-\ref{nonlin}-(a)
\begin{eqnarray}\label{jacobi1}
\frac{dg_1}{dz}=\pm\sqrt{(\lambda_2+\lambda_3g_1^2)(g_1^2-1)}
\end{eqnarray} 
Similar consideration with equation-\ref{nonlin}-(a) and \ref{nonlin}-(c)
\begin{eqnarray}\label{jacobi3}
\frac{dg_3}{dz}=\pm\sqrt{(\lambda_2+\lambda_1g_3^2)(1-g_3^2)}
\end{eqnarray}

In order to express $\{g_1, g_2, g_3\}$ in terms of Jacobi elliptic functions, we impose the constraint $\lambda_1+\lambda_2+\lambda_3=0$ which is maintained throughout section-\ref{witt}. Motivated from the discussion of bi-orthogonality a possible choice of solutions can be $\lambda_1=\gamma^2, \lambda_2=\omega^2, \lambda_3=-1$. Choosing positive signs for the right hand sides of the equations-\ref{jacobi2}, \ref{jacobi1}, \ref{jacobi3}  one can write

\begin{eqnarray}
{\rm{(a)}}\:\:\frac{dg_1}{dz}&=&\sqrt{(1-g_1^2)(g_1^2-\gamma^2+1)}\nonumber\\
{\rm{(b)}}\:\:\frac{dg_2}{dz}&=&\sqrt{(1-g_2^2)(1-\gamma^2g_2^2)}\nonumber\\
{\rm{(c)}}\:\:\frac{dg_3}{dz}&=&\sqrt{(1-g_3^2)(\gamma^2 g_3^2+1-\gamma^2)}
\end{eqnarray}

One can safely consider $\gamma$ and $\omega$ as the modulus $(k)$ and complimentary modulus $(k^\prime)$ respectively for the said elliptic functions $\{g_1, g_2, g_3\}$.

Integrating the above equations under suitable limits, we get $g_1=-dn(z; k), g_2=sn(z; k)$ and $g_3=-cn(z; k)$. It is easy to verify from the equations-\ref{lamb1}, \ref{lamb2}, \ref{lamb3} standard result
\begin{equation}\label{stan1}
sn^2z+cn^2z=1=dn^2z+\gamma^2sn^2z
\end{equation}
and from equation-\ref{nonlin}
\begin{equation}\label{stan2}
\frac{d}{dz}snz=cnzdnz\:\:\frac{d}{dz}cnz=-snzdnz\:\:\frac{d}{dz}dnz=-k^2cnzsnz
\end{equation}

Using the above results (equations-\ref{stan1}, \ref{stan2}) and recalling section-\ref{biortho} the following isomorphism corresponding to CK-algebra of \textbf{elliptic} type is obvious 
\begin{equation}\label{jac1}
\{\widetilde{J}, \widetilde{P}_1, \widetilde{P}_2\}\cong\left\{\frac{i}{\omega}\sigma_1^\gamma, i\sigma_2^\gamma, \frac{-i}{\omega}\sigma_3^\gamma\right\}\cong\left\{-\frac{dn(z; \gamma)}{\omega}\frac{d}{dz}, i\frac{cn(z; \gamma)}{\omega}\frac{d}{dz},isn(z; \gamma)\frac{d}{dz}\right\}.
\end{equation}

\begin{remark}
The above method may also be viewed as a new method of constructing three basic elliptic functions of Jacobi.
\end{remark}

The following choices are possible
\begin{equation}\label{jac2}
\{\widetilde{J}, \widetilde{P}_1, \widetilde{P}_2\}\equiv\left\{-\frac{dn(z; \gamma)}{\omega}\frac{d}{dz}, \frac{cn(z; \gamma)}{\omega}\frac{d}{dz},sn(z; \gamma)\frac{d}{dz}\right\}
\end{equation}

to get the commutations

\begin{equation}
[\widetilde{J}, \widetilde{P}_1]=P_2,\:\:[\widetilde{J}, \widetilde{P}_2]=-\widetilde{P}_1\:\:[\widetilde{P}_1, \widetilde{P}_2]=-\widetilde{J}
\end{equation}

of \textbf{hyperbolic algebras}.

Choice of
\begin{equation}\label{jac3}
\{\widetilde{J}, \widetilde{P}_1, \widetilde{P}_2\}\equiv\left\{\frac{cn(z; \gamma)}{\omega}\frac{d}{dz}, \frac{dn(z; \gamma)}{\omega}\frac{d}{dz},sn(z; \gamma)\frac{d}{dz}\right\}
\end{equation}
leads to \textbf{co-hyperbolic algebra}. On the other hand the choice 
\begin{equation}\label{jac4}
\{\widetilde{J}, \widetilde{P}_1, \widetilde{P}_2\}\equiv\left\{{sn(z; \gamma)}\frac{d}{dz}, -\frac{idn(z; \gamma)}{\omega}\frac{d}{dz},\frac{icn(z; \gamma)}{\omega}\frac{d}{dz}\right\}
\end{equation}
gives \textbf{doubly hyperbolic algebra}.

\subsection{Realizations by Reciprocals and Quotients of Jacobi Functions}\label{witt2}
In terms of \textbf{Glaisher's notation} one may write the following
\begin{eqnarray}
ns(z; k)=\frac{1}{sn(z; k)},\:\:\:nc(z; k)=\frac{1}{cn(z; k)},\:\:\:nd(z; k)=\frac{1}{dn(z; k)}\nonumber\\
sc(z; k)=\frac{sn(z; k)}{cn(z; k)},\:\:\:sd(z; k)=\frac{sn(z; k)}{dn(z; k)},\:\:\:cd(z; k)=\frac{cn(z; k)}{dn(z; k)}\nonumber\\
cs(z; k)=\frac{cn(z; k)}{sn(z; k)},\:\:\:ds(z; k)=\frac{dn(z; k)}{sn(z; k)},\:\:\:dc(z; k)=\frac{dn(z; k)}{cn(z; k)}
\end{eqnarray}

In each of the following tables we consider a particular choice of $\{\lambda_j : j=1, 2, 3\}$ consistent with the constraint $\lambda_1+\lambda_2+\lambda_3=0$.

\vspace{0.5cm}

\textbf{Case-1}: Considering $f_1=\frac{g_1}{\sqrt{\lambda_2}}, f_2=ig_2, f_3=i\frac{g_3}{\sqrt{\lambda_2}}$ and letting $\lambda_1=-\omega^2, \lambda_2=-\gamma^2, \lambda_3=1$ we get $\{g_1, g_2, g_3\}=\{dc(z; \gamma),sc(z; \gamma), nc(z; \gamma)\}$ following the method in section-\ref{witt1}.

Applying the following results from differentiation
\begin{eqnarray}
\frac{d}{dz}ncz=sczdcz,\:\:
\frac{d}{dz}scz=nczdcz,\:\:
\frac{d}{dz}dcz=k^{\prime^2}sczncz
\end{eqnarray}
we derive the \textbf{elliptic algebra} in table-\ref{ta1}. The rest are obtained similarly.

\begin{table}[h]
\caption{Realizations involving $\{nc, sc, dc\}$}
\label{ta1}
	{\begin{tabular}{|c|c|c|c|} \toprule
			Type of CK algebra & $\widetilde{J}$ & $\widetilde{P}_1$ & $\widetilde{P}_2$\\\hline
			Elliptic & $\frac{idc(z; \gamma)}{\gamma}\frac{d}{dz}$ & $\frac{nc(z; \gamma)}{\gamma}\frac{d}{dz}$ & $isc(z; \gamma)\frac{d}{dz}$\\\hline
			Hyperbolic & $\frac{idc(z; \gamma)}{\gamma}\frac{d}{dz}$ & $\frac{-sc(z; \gamma)}{\gamma}\frac{d}{dz}$ & $-inc(z; \gamma)\frac{d}{dz}$\\\hline
			Co-Hyperbolic & $\frac{dc(z; \gamma)}{\gamma}\frac{d}{dz}$ & $\frac{nc(z; \gamma)}{\gamma}\frac{d}{dz}$ & $sc(z; \gamma)\frac{d}{dz}$\\\hline
			Doubly-Hyperbolic & $\frac{dc(z; \gamma)}{\gamma}\frac{d}{dz}$ & $\frac{inc(z; \gamma)}{\gamma}\frac{d}{dz}$ & $isc(z; \gamma)\frac{d}{dz}$\\\hline
		\end{tabular} }
\end{table}

\vspace{0.5cm}

\textbf{Case-2} :
Similarly, considering $f_1=\frac{g_1}{\sqrt{\lambda_3}}, f_2=\frac{g_2}{\sqrt{\lambda_1\lambda_2}}, f_3=\frac{g_3}{\sqrt{\lambda_1}}$ and letting $\lambda_1=\omega^2, \lambda_2=-1, \lambda_3=\gamma^2$ we get $\{g_1, g_2, g_3\}=\{ns(z; \gamma),ds(z; \gamma), cs(z; \gamma)\}$  which constitute the generators of \textbf{elliptic algebra} 
upon using the following results from differentiation
\begin{eqnarray}
\frac{d}{dz}nsz=-cszdsz,\:\:
\frac{d}{dz}csz=-nszdsz,\:\:
\frac{d}{dz}dsz=-csznsz
\end{eqnarray}

Finally  the following table-\ref{ta2} suggests realizations of all the other possible algebras.

\begin{table}[h]
\caption{Realizations involving $\{ns, cs, ds\}$}
\label{ta2}
	{\begin{tabular}{|c|c|c|c|} \toprule
			Type of CK algebra & $\widetilde{J}$ & $\widetilde{P}_1$ & $\widetilde{P}_2$\\\hline
			Elliptic & $\frac{ins(z; \gamma)}{\gamma}\frac{d}{dz}$ & $\frac{ics(z; \gamma)}{\omega}\frac{d}{dz}$ & $\frac{ds(z; \gamma)}{\gamma\omega}\frac{d}{dz}$\\\hline
			Hyperbolic & $\frac{ds(z; \gamma)}{\gamma\omega}\frac{d}{dz}$ & $\frac{ns(z; \gamma)}{\gamma}\frac{d}{dz}$ & $\frac{-ics(z; \gamma)}{\gamma}\frac{d}{dz}$\\\hline
			Co-Hyperbolic & $\frac{cs(z; \gamma)}{\omega}\frac{d}{dz}$ & $\frac{ds(z; \gamma)}{\gamma\omega}\frac{d}{dz}$ & $\frac{ns(z; \gamma)}{\gamma}\frac{d}{dz}$\\\hline
			Doubly-Hyperbolic & $\frac{ns(z; \gamma)}{\gamma}\frac{d}{dz}$ & $\frac{cs(z; \gamma)}{\omega}\frac{d}{dz}$ & $\frac{ds(z; \gamma)}{\gamma\omega}\frac{d}{dz}$\\\hline
		\end{tabular} }
\end{table}

\begin{remark}
The realization of Table-\ref{ta2} is of particular interest. If the normalization factor in each of the generators is withdrawn, one can verify the exact form of un-normalized  CK-algebras with which we have started in equation-\ref{CK}. The curvatures are $(\gamma^2, \omega^2)$ which are also the modulus and complementary modulus respectively of the relevant elliptic functions. For example, the set $\{J, P_1, P_2\}\equiv\left\{ins(z; k)\frac{d}{dz}, ics(z; k)\frac{d}{dz}, ds(z; k)\frac{d}{dz} \right\}$ represents \textbf{elliptic} CK-algebra with fractional curvatures $(\gamma^2, \omega^2)$.  
\end{remark}

\vspace{0.5cm}

\textbf{Case-3} :
Following  the same procedure, by considering $f_1=-{g_1}, f_2=i{g_2}, f_3=i{g_3}$ and letting $\lambda_1=-\gamma^2, \lambda_2=1, \lambda_3=-\omega^2$ we get $\{g_1, g_2, g_3\}=\{nd(z; \gamma),sd(z; \gamma), cd(z; \gamma)\}$  which constitute the generators of \textbf{elliptic algebra} 
upon using the following results from differentiation

\begin{eqnarray}
\frac{d}{dz}ndz=k^2cdzsdz,\:\:
\frac{d}{dz}cdz=-k^{\prime^2}ndzsdz,\:\:
\frac{d}{dz}sdz=cdzndz
\end{eqnarray}

Finally, table-\ref{ta3} is obtained.

\begin{table}[h]
\caption{Realizations involving $\{nd, cd, sd\}$}
		\label{ta3}
	\begin{tabular}{|c|c|c|c|} \toprule
			Type of CK algebra & $\widetilde{J}$ & $\widetilde{P}_1$ & $\widetilde{P}_2$\\\hline
			Elliptic & ${-nd(z; \gamma)}\frac{d}{dz}$ & ${icd(z; \gamma)}\frac{d}{dz}$ & ${isd(z; \gamma)}\frac{d}{dz}$\\\hline
			Hyperbolic & ${nd(z; \gamma)}\frac{d}{dz}$ & ${sd(z; \gamma)}\frac{d}{dz}$ & ${cd(z; \gamma)}\frac{d}{dz}$\\\hline
			Co-Hyperbolic & ${ind(z; \gamma)}\frac{d}{dz}$ & ${isd(z; \gamma)}\frac{d}{dz}$ & ${-cd(z; \gamma)}\frac{d}{dz}$\\\hline
			Doubly-Hyperbolic & ${sd(z; \gamma)}\frac{d}{dz}$ & ${-nd(z; \gamma)}\frac{d}{dz}$ & ${cd(z; \gamma)}\frac{d}{dz}$\\\hline
		\end{tabular} 
		
\end{table}

\begin{remark}
 Each of the said realizations has a direct connection to bi-orthogonal system in $\mathbb{C}^2$. In so far as the deformation parameter $\gamma$ satisfies $\vert\gamma\vert<1$, the said bi-orthogonal systems at all exists. Since $\gamma$ comes out to be the modulus of the relevant elliptic functions it may be said that the condition of bi-orthogonality necessitates the use of elliptic functions.
\end{remark}

\begin{remark}
The above discussion shows that the values of $\{\lambda_1, \lambda_2, \lambda_3\}=\{\pm\gamma^2, \pm\omega^2, \mp 1\}$ have been permuted to obtain various cases in point. For example, if we consider a vector $\mathbf{\Omega}=(\gamma^2\:\:\omega^2\:\: -1)$, case-1, 3, 2 can be obtained by applying
 $\left(\begin{array}{ccc}0 & 1 & 0\\
1 & 0 &0\\
0 &0 & 1
\end{array}\right)$, $\left(\begin{array}{ccc}1 & 0 & 0\\
0 & 0 &1\\
0 &1 & 0
\end{array}\right)$ on $-\Omega$ and $\left(\begin{array}{ccc}0 & 1 & 0\\
0 & 0 &1\\
0 &1 & 0
\end{array}\right)$ on $\Omega$. In fact, keeping $\gamma^2$ and $\omega^2$ with same sign and $1$ with opposite sign twelve such cases are available. Though, we have considered only four cases the remaining options can be reduced to these cases with suitable choice of $\{f_j : j=1, 2, 3\}$ and permutation matrices.
\end{remark}

\subsection{Realizations related to Modular Transformation}\label{mod}
The basic theories on \textbf{Modular Group} and \textbf{Modular Transformation} has been discussed in the \textbf{Appendix}. Let us consider the cases corresponding to two fundamental modular transformations : (i) $\tau=-1/\tau$ and (ii) $\tau^\prime=1+\tau$.

\subsubsection{Realizations by elliptic functions of modulus $k^{\prime}$}
Considering the transformation $\tau^\prime=-1/\tau$ and invoking Jacobi's $\Theta$-functions $\Theta_3^2(0\vert\tau^\prime)=-i\tau\Theta^2_3(0\vert\tau)$. Now by definition
\begin{equation}
sn(u\vert\tau^\prime)=\frac{\Theta_3(0\vert\tau^\prime)}{\Theta_2(0\vert\tau^\prime)}\frac{\Theta_1(z^{\prime}\vert\tau^\prime)}{\Theta_4(z^\prime\vert\tau^\prime)}=-i\frac{\Theta_3(0\vert\tau^\prime)}{\Theta_4(0\vert\tau^\prime)}\frac{\Theta_1(\tau z^\prime\vert\tau)}{\Theta_2(\tau z^\prime\vert\tau)}
\end{equation}
Where, $z^\prime=u\Theta_3^{-2}(0\vert\tau^\prime)$. Now,
\begin{equation}
\tau z^\prime=u\Theta_3^{-2}(0\vert\tau^\prime)=iu\Theta_3^{-2}(0\vert\tau^\prime)=iz
\end{equation} 
This gives
\begin{equation}
sn(u\vert\tau^\prime)=-i\frac{\Theta_3(0\vert\tau^\prime)}{\Theta_4(0\vert\tau^\prime)}\frac{\Theta_1(\tau z^\prime\vert\tau)}{\Theta_2(\tau  z^\prime\vert\tau)}=-isc(iu\vert\tau)
\end{equation}
Or, $sn(u;k^\prime)=-isc(iu, k)$
Similar realizations involving modulus $k^{\prime}$  can be possible by using the following table with $\zeta=iz$ (see \textbf{Appendix}) and the discussion stated above.

\begin{table}[h]\label{ta4}
\caption{Realizations involving interchangeability of $k$ and $k^\prime$}
	{\begin{tabular}{|c|c|c|} \toprule
			$sc(z; k^{\prime})=-isn(\zeta; k)$ & $nc(z; k^{\prime})=cn(\zeta; k)$ & $dc(z; k^{\prime})=dn(\zeta; k)$ \\\hline
			$cs(z; k^{\prime})=ins(\zeta; k)$ & $ds(z; k^{\prime})=ids(\zeta; k)$ & $ns(z; k^{\prime})=ics(\zeta; k)$ \\\hline
			$nd(z; k^{\prime})=cd(\zeta; k)$ & $cd(z; k^{\prime})=nd(\zeta; k)$ & $sd(z; k^{\prime})=-isd(\zeta; k)$ \\\hline
			$cn(z; k^{\prime})=nc(\zeta; k)$ & $sn(z; k^{\prime})=-isc(\zeta; k)$ & $dn(z; k^{\prime})=dc(\zeta; k)$ \\\hline
		\end{tabular} }
\end{table}
For example from Table-\ref{ta1} the generators of elliptic algebra are given by $\left\{-\frac{1}{k^{\prime}}dn(\zeta;k)\frac{d}{d\zeta}, i\frac{1}{k^{\prime}}cn(\zeta;k)\frac{d}{d\zeta}, isn(\zeta;k)\frac{d}{d\zeta}\right\}$.

\subsubsection{Realizations by Jacobi functions of arbitrary absolute value of the modulus $\vert\lambda\vert\in (0, \infty)$}

Considering the modular transformation $\tau^\prime=1+\tau$, the new nome $q^\prime=e^{i\pi(1+\tau)}=-q$ 
\begin{equation}
\Theta_1(u\vert\tau^\prime)=\sqrt{i}\Theta_1(u\vert\tau),\:\:
\Theta_2(u\vert\tau^\prime)=\sqrt{i}\Theta_2(u\vert\tau),\:\:
\Theta_3(u\vert\tau^\prime)=\Theta_4(u\vert\tau),\:\:
\Theta_4(u\vert\tau^\prime)=\Theta_3(u\vert\tau)
\end{equation}
Hence by definition the new moduli are 
\begin{eqnarray}
\lambda=\left(\frac{\Theta_2(0\vert\tau^\prime)}{\Theta_3(0\vert\tau^\prime)}\right)^2=i\left(\frac{\Theta_2(0\vert\tau)}{\Theta_4(0\vert\tau)}\right)^2=\frac{ik}{k^\prime}\nonumber\\
\lambda^\prime=\left(\frac{\Theta_4(0\vert\tau^\prime)}{\Theta_3(0\vert\tau^\prime)}\right)^2=\left(\frac{\Theta_3(0\vert\tau)}{\Theta_4(0\vert\tau)}\right)^2=\frac{1}{k^\prime}
\end{eqnarray}
\vspace{0.5cm}

Since, $\vert\lambda\vert=\frac{k}{\sqrt{1-k^2}}$ and $k\in (0, 1)$, $\vert\lambda\vert\in (0, \infty)$. Using $z=u/\Theta^2_3(0\vert\tau)$ and $z^\prime=u/\Theta^2_3(0\vert\tau^\prime)$ and hence $z^\prime=z/k^\prime$.
One can therefore write following the definition of $sn(u; \lambda^\prime)$
\begin{equation}
sn(u; \lambda)=\frac{\Theta_3(0\vert\tau^\prime)}{\Theta_2(0\vert\tau^\prime)}\frac{\Theta_1(z^\prime\vert\tau^\prime)}{\Theta_4(z^\prime\vert\tau^\prime)}=\frac{\Theta_4(0\vert\tau^\prime)}{\Theta_2(0\vert\tau^\prime)}\frac{\Theta_1(z^\prime\vert\tau^\prime)}{\Theta_3(z^\prime\vert\tau^\prime)}=k^\prime sd(u/k^\prime; k)
\end{equation}

Similarly, $cn(u; \lambda)=cd(u/k^\prime, k)$ and $dn(u; \lambda)=nd(u/k^\prime; k)$.

\begin{eqnarray}
sd(z; k)=\frac{1}{k^{\prime}}sn(k^{\prime}z ; \lambda),\:\: cd(z; k)=cn(k^{\prime}z ; \lambda),\:\: nd(z; k)=dn(k^{\prime}z ; \lambda)
\end{eqnarray}
where, $u=k^{\prime}z$ and $\lambda=ik/k^{\prime}$.
Now using Table-\ref{ta3} one can write the representations of all possible CK-algebras.  For example, the generators of elliptic algebra are given by $\left\{-k^{\prime}dn(u;\lambda)\frac{d}{du}, ik^{\prime}cn(u;\lambda)\frac{d}{du}, isn(u;\lambda)\frac{d}{du}\right\}$.

\vspace{0.5cm}

\section{Discussion on intrinsic Casimir operator}\label{casimir}
It is interesting to note that the quadratic Casimir Operator as defined in section-\ref{ck} for a given algebra produces different results depending upon the representation considered. In fact, it seems quite perplexing that while Casimirs for the $2\times 2$ matrix representation of any of the algebras is giving non-vanishing expression, the representation by elliptic functions leads to zero values of the same challenging the status of the Casimir as an invariant. In order to resolve the impasse it may be suggestive to introduce a kind of \textbf{Intrinsic Casimir Operator (ICO)} which is representation-independent and derivable purely from from Lie bracket relations with no a priori definition of the Lie bracket itself. The invariance of the Casimir has been argued in view of vanishing Lie derivative \cite{thomas06}. Following \cite{thomas06} let us define the so called ICO of the elliptic algebra
\begin{equation}
\mathcal{C}_E=[\tilde{J}, [\tilde{J}, \cdot]]+[\tilde{P}_1, [\tilde{P}_1, \cdot]]+[\tilde{P}_2, [\tilde{P}_2, \cdot]]
\end{equation}

It can readily be verified that any element $A$ of the same algebra given by $A=aJ+bP_1+cP_2$ under the action of $\mathcal{C}_E$ leads to
\begin{equation}
\mathcal{C}_E(A)=[\tilde{J}, [\tilde{J}, A]]+[\tilde{P}_1, [\tilde{P}_1, A]]+[\tilde{P}_2, [\tilde{P}_2, A]]=2A
\end{equation}
Formally, $\mathcal{C}_E\dot{=}\left(\left[\tilde{J}\right.\right)^2+\left(\left[\tilde{P}_1\right.\right)^2+\left(\left[\tilde{P}_2\right.\right)^2$. Similar Casimir operators can be constructed for all the other algebras. In the case anti-de-Sitter algebra $\mathcal{C}_{AdS}\dot{=}\left(\left[\tilde{J}\right.\right)^2-\left(\left[\tilde{P}_1\right.\right)^2+\left(\left[\tilde{P}_2\right.\right)^2$.

\section{Realizations at the limiting values of the moduli}
Letting the modulus $k$ and complimentary modulus $k^{\prime}$ to tend to limits we sometimes recover vector fields involving trigonometric and hyperbolic functions. However, for CK-algebras not all vector fields are reducible to non-trivial limits.    The following discussion categorizes the vector fields accordingly.

\textbf{Case-1 : }
 $\gamma(=k)\rightarrow 0$ and hence $\omega (=k^{\prime})\rightarrow 1$ : equation-\ref{jac1}  gives
\begin{eqnarray}
\left\{-\frac{d}{dz}, i\cos z\frac{d}{dz}, i\sin z\frac{d}{dz}\right\}
\end{eqnarray}
and similar expressions corresponding to equation-\ref{jac2}, \ref{jac3}, \ref{jac4} and for vector fields of Table-\ref{ta3}. Such limits do not give anything meaningful for Table-\ref{ta1}, \ref{ta2} in the present context.

\textbf{Case-2 : } $\gamma(=k)\rightarrow 1$ and hence $\omega (=k^{\prime})\rightarrow 0$ : This case is prohibited for equation-\ref{jac1}, \ref{jac2}, \ref{jac3}, \ref{jac4}  and for Table-\ref{ta2} however, this works well with Table-\ref{ta1} and Table-\ref{ta3}. Now in view of Table-\ref{ta3}, one can write 
\begin{eqnarray}
\lim_{k\rightarrow 1}sn(z; k)=\lim_{k^\prime\rightarrow 1}cn(z; k^\prime)=\lim_{k\rightarrow 0}-isc(iz; k)=-i\frac{\sin iz}{\cos iz}=\tanh z\nonumber\\
\lim_{k\rightarrow 1}cn(z; k)=\lim_{k^\prime\rightarrow 1}cn(z; k^\prime)=\lim_{k\rightarrow 0}nc(iz; k)=\frac{1}{\cos iz}=\sech z\nonumber\\
\lim_{k\rightarrow 1}dn(z; k)=\lim_{k^\prime\rightarrow 1}dn(z; k^\prime)=\lim_{k\rightarrow 0}dc(iz; k)=\frac{1}{\cos iz}=\sech z 
\end{eqnarray} 
Therefore, in Table-\ref{ta1}
\begin{eqnarray}
\lim_{k\rightarrow 1}dc(z, k)=1,\:\:\lim_{k\rightarrow 1}nc(z, k)=\cosh (z; k),\:\:\lim_{k\rightarrow 1}sc(z, k)=\sinh (z; k)
\end{eqnarray}
Hence the elliptic CK-algebra has the generator $\left\{i\frac{d}{dz}, \cosh (z; k)\frac{d}{dz}, i\sinh (z; k)\frac{d}{dz}\right \}$. Similar method 
 can be employed for Table-\ref{ta3} as well.

\begin{remark}
It is to be noted that vector fields in Table-\ref{ta2} are valid strictly for elliptic functions since neither of the limiting procedures is applicable for them.
\end{remark}

\section{Comments}
The above discussion relates elliptic functions of various kinds with the realization of 2-D Cayley-Klein algebras having non-vanishing curvatures. A triplet of non-linear coupled ordinary differential equations facilitates the determination of the relevant vector fields. Compatibility of these equations with the commutator brackets leads us to elliptic functions of Jacobi.  Two important observations must be mentioned regarding the above realizations.

(i) The Lie bracket relations specific to CK-algebra considered above may be viewed as identities of elliptic functions and atleast for one case there exists no analogue in terms of trigonometric and hyperbolic limits i. e.; the respective generators non-trivially involve elliptic functions.

(ii) The moduli of the elliptic functions and the so called modular transformations play a crucial role in different realizations.

The method seems to have appreciable significance when exploited to extend it for higher dimension which requires suitable choices of coupled differential equations in order to land upon new type of functions as their solutions. Some of the results regarding this possibility are under preparation.

\section*{Appendix : Some results on elliptic functions}

\textbf{Definitions of Jacobi Elliptic Functions} : 
Jacobi elliptic functions can be defined with the help of Jacobi theta functions $\{\Theta_j(u\vert \tau) ; j=1,\cdots, 4\}$ defined as follows
\begin{eqnarray}
\Theta_1(u\vert \tau)&=&2\sum_{n=0}^\infty(-1)^nq^{(n+\frac{1}{2})^2}\sin(2n+1)u\nonumber\\
\Theta_2(u\vert \tau)&=&2\sum_{n=0}^\infty q^{(n+\frac{1}{2})^2}\cos(2n+1)u\nonumber\\
\Theta_1(u\vert\tau)&=&1+2\sum_{n=1}^\infty q^{n^2}\cos 2nu\nonumber\\
\Theta_4(u\vert\tau)&=&1+2\sum_{n=1}^\infty(-1)^nq^{n^2}\cos 2nu\:\:\forall\:\:\vert q\vert<1.
\end{eqnarray}
considering $z=u\Theta^2_3(0\vert\tau)$, the functions $\{sn(z; k), cn(z; k), dn(z; k)\}$ are defined as
\begin{equation}
sn(z; k)=\frac{\Theta_3(0\vert\tau)}{\Theta_2(0\vert\tau)}\frac{\Theta_1(u\vert\tau)}{\Theta_4(u\vert\tau)},\:\:
cn(z; k)=\frac{\Theta_4(0\vert\tau)}{\Theta_2(0\vert\tau)}\frac{\Theta_2(u\vert\tau)}{\Theta_4(u\vert\tau)},\:\:
dn(z; k)=\frac{\Theta_4(0\vert\tau)}{\Theta_3(0\vert\tau)}\frac{\Theta_3(u\vert\tau)}{\Theta_4(u\vert\tau)}.
\end{equation}
where, the \textbf{modulus} $k=\Theta_2^2(0\vert\tau)/\Theta_3^2(0\vert\tau)$ and \textbf{complementary modulus} $k^{\prime}=\Theta_4^2(0\vert\tau)/\Theta_3^2(0\vert\tau)$. It is easy to verify that $k^2+k^{{\prime}^2}=1$. $q=e^{i\pi\tau}$ is called the \textbf{nome}. Since the imaginary part of $\tau$ is considered to be positive $\vert q\vert<1$. In terms of $q$, $k$ and $k^{\prime}$ have the expressions 
\begin{equation}
k=4\sqrt{q}\prod_{n=1}^\infty\left(\frac{1+q^{2n}}{1+q^{2n-1}}\right)^4\:\:{\rm{and}}\:\:k^{\prime}=\prod_{n=1}^\infty\left(\frac{1-q^{2n-1}}{1+q^{2n-1}}\right)^4
\end{equation}

\textbf{Modular Transformation} : Modular transformation is given by
\begin{equation}
\tau\rightarrow\tau^\prime=\frac{c+d\tau}{a+b\tau};\:\: a, b, c, d\in\mathbb{Z}^\pm\cup\{0\}\:\:{\rm{and}}\:\:ad-bc=1 
\end{equation}
It has been found that all possible modular transformations can be represented as product of two fundamental types $\tau^\prime=1+\tau$ and $\tau^\prime=-1/\tau$ whose corresponding matrix-representations are given by
\begin{equation}
P=\left(\begin{array}{cc}
	1 & 0 \\
	1 & 1 
\end{array} \right),\:\:\:Q =\left(\begin{array}{cc}
	0 & 1 \\
	-1 & 0 
\end{array} \right)
\end{equation}

$\{P, Q\}$ are generators of the so called \textbf{Modular Group}\cite{lawden89} whose general element is given by $M=\left(\begin{array}{cc}
	a & b \\
	c & d 
\end{array} \right)$.

\section*{Acknowledgement} AC wishes to thank his colleague Dr. Baisakhi Mal for extending her support in manuscript preparation.

\vspace{0.5cm}

\textbf{Funding Declaration} : No funding has been received by the author from any funding agency or institute.

\vspace{0.5cm}

\textbf{Competing Interest declaration} :  The author declares that he has no significant competing financial, professional, or personal interests that might have influenced the performance or presentation of the work described in this manuscript. 

\vspace{0.5cm}

\textbf{Author Contribution declaration} : The author declares that he is the only author of the manuscript and hence responsible for the preparation of the entire manuscript.

\vspace{0.5cm}

ORCID iDs

Arindam Chakraborty https://orcid.org/0000-0002-3414-3785

\end{document}